%% file: transfernew2.tex
\documentclass{svmult}

\usepackage{makeidx}         
\usepackage{graphicx}        
\usepackage{multicol}        
\usepackage[bottom]{footmisc}

\makeindex             

\begin{document}

\title*{Prospects for Money Transfer Models}
\author{Yougui Wang\inst{*}\and
Ning Ding\and Ning Xi}
\institute{Department of Systems Science, School of Management,
Beijing Normal University, Beijing, 100875, People's Republic of
China \texttt{ygwang@bnu.edu.cn}}
%
%
\maketitle
\begin{abstract}
Recently, in order to explore the mechanism behind wealth or
income distribution, several models have been proposed by applying
principles of statistical mechanics. These models share some
characteristics, such as consisting of a group of individual
agents, a pile of money and a specific trading rule. Whatever the
trading rule is, the most noteworthy fact is that money is always
transferred from one agent to another in the transferring process.
So we call them money transfer models. Besides explaining income
and wealth distributions, money transfer models can also be
applied to other disciplines. In this paper we summarize these
areas as statistical distribution, economic mobility, transfer
rate and money creation. First, money distribution (or income
distribution) can be exhibited by recording the money stock
(flow). Second, the economic mobility can be shown by tracing the
change in wealth or income over time for each agent. Third, the
transfer rate of money and its determinants can be analyzed by
tracing the transferring process of each one unit of money.
Finally, money creation process can also be investigated by
permitting agents go into debts. Some future extensions to these
models are anticipated to be structural improvement and
generalized mathematical analysis.

\end{abstract}

\section{Introduction}
\label{sec:1}
Money does matter to an economy. To understand the role that money
plays in the performance of economic system, many theoretical
studies have been performed in traditional economics. Recently, a
small branch of ``econophysicists" shifted their attentions to
this issue. Several models have been developed by applying
principles of statistical mechanics to the questions of income and
wealth distribution \cite{model1, model2, model3, model4, follow}.
These models share some characteristics, such as consisting of a
group of individual agents, a pile of money and a specific trading
rule. The most noteworthy fact is that money is always transferred
from one agent to another in the transferring process. So this
kind of models could be referred as money transfer models. The
prime theme of constructing such models is to explore the
mechanism behind wealth or income distribution. In fact, they can
be applied more widely in some other economic issues. In this
paper, we prospect for some applications of these transfer models
and anticipate that considerable achievements can be made on the
basis of them. We also argue that further improvements should be
accomplished to make these models much more realistic.

The purpose of this paper is to identify what issues could be
analyzed on the basis of money transfer models. This kind of
models is very easy to grasp, for only two elements are involved:
money and agents. Money is possessed or held by agents, and may be
transferred among them via trading. Based on these models, recent
efforts were mainly devoted to the formation of monetary wealth
distribution, the circulation of money \cite{wang,ding} and
creation of money \cite{Braun3}. We would like to summarize and
expand the scope of their applications in the following four
routes.

\section{Applications}
\label{sec:2}
\subsection{Distribution}\label{sec:2.1}
Money transfer models are originally used to demonstrate steady
distributions of money. This can be achieved by recording the
quantity of money stock possessed by each agent in the
simulations. In the basic model proposed by A. Dr\u{a}gulescu and
V.M. Yakovenko, the money distribution follows a Boltzmann-Gibbs
law \cite{model1}. B.K. Chakrabarti et al. introduced the saving
behavior into the model \cite{model2,model3}, and found the money
distribution obeys a Gamma law when all the agents are set with
the same saving factor, but a power law as the saving factor is
set randomly. N. Ding et al. introduced the preferential
dispensing behavior into the trading process and also obtained a
stationary power-law distribution \cite{model4}. From these
results we can see that the shape of distribution is determined by
the trading rule.

Besides these theoretical studies, econophysicists also performed
the empirical studies on the distribution in the economy,
following the earlier Pareto's work. The analysis showed that in
many countries the income distribution typically presents with a
power-law tail, and majority of the income distribution can be
described by an exponential distribution \cite{emp1,emp2,emp3}. It
is worthy noting that account of these empirical studies is taken
of income distribution. Income corresponds to money flow which is
different from money amount. However, all the distributions
presented in previous simulations do not refer to the money flow.
Actually, in the money transferring process, we can also record
the level of money flow received by each agent during a given
period. The statistics of them yields the flow type distribution.
Thus, embodying the money flow generation mechanism, the transfer
models can also provide a convenient tool for investigating the
mechanism behind the income distribution in reality.

\subsection{Mobility}\label{sec:2.2}
During the simulations of money transfer models, the amount of
money held by agents varies over time. This phenomenon is called
mobility in economics. In the view of economists, mobility is an
indispensable supplement to distribution because the former can
cure the anonymity assumption of the latter \cite{mob}. And the
analysis of mobility is greatly helpful to comprehend the dynamic
mechanism behind the distribution. In addition, like distribution,
economic mobility should be an essential criterion when evaluating
a relevant theoretical model.

In the transferring process, the economy will reach its steady
state and the distribution will keep unchanged. After that, the
amount of money still fluctuates over time for each agent,
meanwhile the rank of each agent shifts from one position to
another. To show the mobility phenomenon with clarity, we can
record agents' rank instead of the amount of money. The time
series of rank for any agent's can be obtained by sorting all of
agents according to their money in the end of each round. We
performed some simulations and the primary results show all of
agents are equal in the economies of models in Ref. \cite{model1}
and \cite{model2}. They have the same probability to be the rich
or the poor. It can be found that the frequency of the rank
fluctuation decreases as the saving rate increases. By contrast,
the economy in Ref. \cite{model3} is stratified where agents are
not equal any longer for their saving rates are set diversely.
Based on these results, it can be concluded that different models
exhibit different mobility characters.

\subsection{Transfer Rate}\label{sec:2.3}
In reality, money does not remain motionless. Instead, it is
transferred from hand to hand consecutively. This phenomenon is
called the circulation of money in economics. The term usually
used to describe the circulation is the velocity of money, which
can be computed by the ratio of total transaction volume to the
money stock. In fact, it refers to the transfer rate of money that
measures how fast the money moves between agents. This rate can be
observed by recording the time intervals for each unit of money to
be held. This kind of time interval is called "holding time" or
"latency time" of money. It can be found that there is not only a
distribution of money among agents, but also a steady distribution
of holding time as the economy reaches its equilibrium state. The
holding time distribution also shifts its shape depending on the
trading rule. For instance, in the simulation of the model with
uniform saving factor the stationary distribution of holding time
obeys exponential law, while in the model with diverse saving
factor the distribution changes to a power type \cite{ding}.

The transfer rate of money has an inverse relation with the
average holding time of money. When the circulation process is in
the nature of Poisson one, the probability distribution of the
latency time of money takes the following form \cite{wang}
\begin{equation}
P(t)=\frac{1}{T}e^{-\frac{t}{T}},
\end{equation}
where $1/T$ corresponds to the intensity of Poisson process, and
$T$ signifies the average holding time of money. In this case, the
velocity of money can be written as
\begin{equation}
V=\frac{1}{T}.
\end{equation}

Since the average holding time is governed by the money holders
(agents in the models), the above equation suggests that the
velocity is determined by the behavior patterns of economic
agents. Employing the well-known life-cycle model in economics,
Wang et al. demonstrated that the velocity of money can be
obtained from the individual's optimal choice \cite{wang1}. Thus
the study on the transferring process provides a new insight into
the velocity of money circulation.

\subsection{Money Creation}\label{sec:2.4}
With the help of money transfer models, we can still discuss the
impact of money creation on the statistical mechanics of money
circulation. In reality, most part of the monetary aggregate that
circulates in the modern economy is created by debts through
banking system. Thus money creation has important influence on the
characteristics of monetary economic system.

Recently, some investigations have been carried out in this line
mainly from two perspectives. One is from physics perspective.
Adrian Dr\u{a}gulescu and Victor Yakovenko demonstrated the
equilibrium probability distribution of money follows the
Boltzmann-Gibbs law, allowing agents to go into debt and putting a
limit on the maximal debt of an agent \cite{model1}. Robert
Fischer and Dieter Braun analyzed the process of creation and
annihilation of money using a mechanical method and examined how
money creation affects statistical mechanics of money
\cite{Braun3}. The other is from economics perspective. It is
known that the essence of money creation can be represented by the
required reserve ratio from the multiplier model of money in
economics. Thus we can examine the dependence of monetary wealth
distribution and the velocity of money on the required reserve
ratio based on a transfer model of money and computer simulations.
We extended a money transfer model by introducing a banking
system, where money creation is achieved by bank loans and the
monetary aggregate is determined by the monetary base and the
required reserve ratio. The simulation results show that monetary
wealth follows asymmetric Laplace distribution, and the velocity
decreases as the required reserve ratio increases. For more
details you can see Ref. \cite{xi}.

\section{Discussion and Conclusion}
\label{sec:3}

The money transfer models were constructed originally for
explaining the real income or wealth distribution. They also can
be applied to other economic issues, such as economic mobility,
transfer rate and money creation. These applications will bring
this kind of models to be rival to the prevailing models in
monetary economics. Of course, the current version of these models
is far from perfectness. In order to fulfill the goal, some
further improvements and modifications are required. One is to
make the agents in the model closer to rational economic ones.
Another one is to analyze the model in a generalized mathematical
way, which would help us to understand the model deeply and
completely and show the right way to structural modification.

\input{referenc}



\printindex
\end{document}

%% file: referenc.tex
%
%

%
%